\documentstyle[12pt]{article}
\begin{document}
\begin{center}
\Large{\bf{No Scalar Hair Theorem for a\\
 Charged Spherical Black Hole}}
\end{center}
\par
\vspace{5mm}
\begin{center}
{\bf{N.Banerjee$^{*}$ and S.Sen$^{**}$}}\\
Relativity and Cosmology Research Centre,\\
Department of Physics,\hspace{2mm}Jadavpur University,\\
Calcutta-700032,\hspace{2mm}India.\\
\end{center}
\vspace{8mm}
\begin{center}
\large{\bf{Abstract}}
\end{center}
\vspace{3mm}
\par
This paper consolidates the no scalar hair theorem for a charged spherically
symmetric black hole in four dimensions in general relativity as well as in all 
scalar tensor theories, both minimally and nonminimally coupled, when the effective 
Newtonian constant of gravity is positive. However, there is an exception when the 
matter field itself is coupled to the scalar field, such as in dilaton gravity.
\par
\vspace{70mm}
\par
$^{*}$narayan@juphys.ernet.in
\par
$^{**}$somasri@juphys.ernet.in
\newpage
\section{Introduction}
\par
Very recently Saa$^{1}$ deduced a theorem which shows that the static spherically
symmetric exterior solutions for the gravitational field equations in a very wide 
class of scalar tensor theories will essentially reduce to the wellknown Schwarzschild 
solutions if one has to hide the essential singularity at the centre of symmetry 
by an event horizon. The no-hair conjecture in general relativity says that 
in the exterior of a black hole, the only information available regarding the 
black hole may be that of its mass, charge and angular momentum. For an uncharged 
spherical black hole, therefore, the only information available to an external 
observer is about its mass. 
Saa$^{1}$ used the solution for the uncharged spherically symmetric mass distribution
along with a minimally coupled scalar field as discussed extensively by Xanthopoulos
and Zannias$^{2}$. It is readily found that the scalar field becomes trivial 
if one demands a regular horizon at a finite distance from the centre of symmetry.
This is in perfect agreement with the no-hair conjecture. The importance of Saa's
work lies in the fact that by virtue of a class of transformations and choice of functions,
he could generate the corresponding static spherically symmetric exterior 
solutions in a large number of scalar tensor theories and thus could show the 
validity of "no scalar hair theorem"in these theories. The scalar field in 
these theories becomes a constant and thus trivial if one demands a blackhole 
i.e, a well defined event horizon shielding the singularity at the centre.
\par
In the case of a spherically symmetric system with an electromagnetic field, the 
same result is expected, i.e, for the existence of an event horizon, the scalar 
field is likely to become trivial. But this conclusion cannot be taken for granted because 
of the nonlinearity of the gravitational field equations. 
\par
In the present work, we make an attempt to generalise Saa's work to include an 
electromagnetic field. We start with the action for a very general scalar tensor 
theory of gravity coupled with a Maxwell field,
$$
S[g_{\mu\nu},\phi,F_{\mu\nu}]=\int[f(\phi)R-h(\phi)g^{\mu\nu}\phi_{,\mu}\phi_{,\nu}
-F_{\mu\nu}F^{\mu\nu}]\sqrt{-g}d^{4}x,
\eqno{(1.1)}$$
where $g_{\mu\nu},\phi$ and $F_{\mu\nu}$ are the metric tensor, the scalar field, 
and the Maxwell field respectively, $f(\phi)$ and $h(\phi)$ are positive 
functions of the scalar field. It should be noted that the scalar field is 
nonminimally coupled to gravity and as a result the Newtonian constant G becomes 
a function of $\phi$. This action in special cases reduces to the corresponding 
actions for Brans-Dicke$^{4}$ theory or the generalized scalar tensor theory of 
Nordtvedt$^{5}$. For Brans-Dicke theory $f(\phi)=\phi$ and $h(\phi)={\omega\over{\phi}}$
where $\omega$ is a constant parameter. If $\omega$ is a function of $\phi$
instead of being constant with the same choice for $f(\phi)$ and $h(\phi)$,
the action corresponds to that for Nordtvedt's generalised scalar tensor theory.
For $f(\phi)=1-{1\over{6}}\phi^{2}$ and $h(\phi)={1\over{2}}$ we get the conformally
coupled scalar field of Bekenstein$^{3}$.
\par
In the present work we show that by using a conformal transformation of the 
metric tensor and redefining the scalar field in a similar way as discussed
by Saa, the nonminimal coupling between the geometry and the scalar field can be 
broken and the action would correspond to that of an Einstein-Maxwell field along with
a zero mass scalar meson field. The corresponding solutions for a static spherically 
symmetric spacetime had been given by Penney$^{6}$ way back in 1969. We show that in Penney's 
solution the scalar field becomes trivial if one demands a regular horizon.
We use the inverse transformation of $g_{\mu\nu}$ to get corresponding solutions in 
a broad class of nonminimally coupled theories in order to investigate the conditions for a 
black hole to exist. The result is the same, the scalar field becomes trivial when we demand
that an event horizon should exist. It deserves mention, however, that the function 
$f(\phi)$ is restricted to positive values only.
\par
In the case of a conformally coupled scalar field, where $f(\phi)=(1-{\phi^{2}\over{6}})$,
Bekenstein$^{3}$ obtained a black hole solution but indicated it is not a genuine 
black hole as the scalar field diverges at the event horizon. In a subsequent paper$^{7}$, however, he proved 
that there is no geodetic incompleteness in the motion of a test particle and the tidal forces 
are all regular even in the nontrivial presence of the scalar field and concluded
that this example indeed represents a genuine black hole. Hence this gives a counter
example to the no scalar hair theorem. The value of $f(\phi)$, in this case is 
negative near the event horizon. Very recently Mayo and Bekenstein$^{8}$
considered the no scalar hair theorem extensively for a multiplet of minimally coupled
scalar fields as well as for a non-minimally coupled scalar field where the coupling with 
the Ricci scalar in the lagrangian is given by
$$
f(\phi)=1-\chi\phi^{2}.$$
They proved the non existence of the scalar hair in such cases for a general value
of $\chi$, namely $\chi<0$ or $\chi>{1\over{2}}$.The work of Mayo and Bekenstein is very
powerful as it takes into account self interacting scalar fields, i.e, nonlinear potential
functions in the action.
\par
The case of a multiplet of scalar fields had, infact, been discussed 
previously by Bekenstein$^{9}$. In this work Bekenstein also extended 
his work to some nonminimally coupled generalised scalar tensor theory$^{5}$ 
for a conformally transformed version of the theory (see Dicke$^{10}$).
\par
Although the present work does not discuss the inclusion of potentials, it widens 
the scalar no hair theorem in the sense that it is valid not only in general relativity
but practically in all scalr tensor theories for $f>0$ with the known exception 
of dilaton gravity.
\par
In next section, the conformal transformation which breaks the nonminimal coupling
between geometry and scalar field is discussed and the corresponding static solutions
for a charged sphere with a minimally coupled scalar field are given. Section 3
deals with some examples of nonminimally coupled scalar tensor theories and in section 4
a brief discussion on the already known results regarding charged black holes in dilaton gravity is given. In section 5
we include a discussion on the results obtained.
\section{Charged spherical black holes with a \newline minimally coupled scalar field}
\par
The coupled Einstein-Maxwell-Scalar field equations, obtained by varying the action 
(1.1) with respect to $g_{\mu\nu}, \phi$ and the Maxwell potential A$^{\mu}$, are given by
$$
f(\phi)R_{\mu\nu}-h(\phi)\phi_{,\mu}\phi_{,\nu}-\{{f(\phi)_{,\mu}}\}_{;\nu}-{1\over{2}}
g_{\mu\nu}\Box f(\phi)=(T_{\mu\nu}-{1\over{2}}Tg_{\mu\nu}),
\eqno{(2.1)}$$
$$
2h(\phi)\Box\phi-h^{\prime}(\phi)g^{\mu\nu}\phi_{,\mu}\phi_{,\nu}+f^{\prime}(\phi)R=0,
\eqno{(2.2)}$$
and
$$
F^{\mu\nu}_{;\nu}=0
\eqno{(2.3)}$$
respectively, where units are so chosen that 8$\pi$G and c are unity. T$_{\mu\nu}$
represents the energy momentum tensor for the electromagnetic field, given by
$$
T_{\mu\nu}=g^{\alpha\beta}F_{\alpha\mu}F_{\beta\nu}-{1\over{4}}g_{\mu\nu}F_{\alpha\beta}F^{\alpha\beta}.
\eqno{(2.4)}$$
For such a field, T, the trace of T$_{\mu\nu}$ is zero. By virtue of the Bianchi 
identity, the wave equation for the scalar field, (2.2), is not an independent equation,
but rather follows from the other field equations.
\par
With a conformal transformation of the form
$$
g_{\mu\nu}=\Omega^{2}\bar{g}_{\mu\nu},
\eqno{(2.5)}$$
the Ricci scalar $R$ transforms as$^{11}$
$$
R=\Omega^{-2}\bar{R}-6\Omega^{-3}\bar{\Box}\Omega.
\eqno{(2.6)}$$
An overhead bar indicates that the variables are in the transformed version. If we choose
$$
\Omega^{-2}=f(\phi),
\eqno{(2.7)}$$
and redefine the scalar field  as $\bar{\phi}$, given by,
$$
\bar{\phi}(\phi)=\sqrt{2}\int^{\phi}_{\phi_{0}}d\xi\sqrt{{3\over{2}}({d\over{d\xi}}
\ln f(\xi))^{2}+{h(\xi)\over{f(\xi)}}},
\eqno{(2.8)}$$
the action (1.1) becomes 
$$
\bar{S}[\bar{g}_{\mu\nu},\bar{\phi},\bar{F}_{\mu\nu}]=
\int[\bar{R}-{1\over{2}}\bar{g}_{\mu\nu}\bar{\phi_{,\mu}}\bar{\phi_{,\nu}}-
\bar{F^{2}}]\sqrt{-\bar{g}}d^{4}x.
\eqno{(2.9)}$$
This result is obtained for any arbitrary positive definite lower limit $"\phi_{0}"$
in equation (2.8). This method is exactly similar to the one discussed by Saa,
the only difference being, in that work the electromagnetic field had not been 
considered. It is easy to check that 
$$
\sqrt{-g}F^{2}=\sqrt{-g}F_{\mu\nu}F^{\mu\nu}=\sqrt{-\bar{g}}\bar{F}_{\mu\nu}
\bar{F}^{\mu\nu}=\sqrt{-\bar{g}}\bar{F^{2}}.
\eqno{(2.10)}$$
\par
The field equations now become formally similar to those for a coupled Einstein-
Maxwell system along with a minimally coupled scalar field.The equations look like
$$
\bar{G}_{\mu\nu}=\bar{\phi}_{,\mu}\bar{\phi}_{,\nu}-{1\over{2}}\bar{g}_{\mu\nu}
\bar{\phi}^{,\alpha}\bar{\phi}_{,\alpha}-\bar{g}^{\alpha\beta}\bar{F}_{\alpha\mu}
\bar{F}_{\beta\nu}+{1\over{4}}\bar{g}_{\mu\nu}\bar{F}_{\alpha\beta}\bar{F}^{\alpha\beta},
\eqno{(2.11)}$$
$$
(\bar{F}^{\mu\nu})_{;\nu}=0,
\eqno{(2.12)}$$
and
$$
\bar{\Box}\bar{\phi}=0.
\eqno{(2.13)}$$
The solutions for this set of equations had been given by Penney$^{6}$.
The line element for the static spherically symmetric spacetime is written as
$$
ds^{2}=-e^{\bar{\gamma}}dt^{2}+e^{\bar{\alpha}}dr^{2}+e^{\bar{\beta}}(d\theta^{2}
+\sin{\theta^{2}}d\phi^{2})
\eqno{(2.14)}$$
In the coordinates where $\bar{\alpha}+\bar{\gamma}=0$, the solutions are
$$
e^{\bar{\alpha}}=e^{-\bar{\gamma}}=(r-a)^{-\Lambda}(r-b)^{-\Lambda}
\left[{b(r-a)^{\Lambda}-a(r-b)^{\Lambda}\over{(b-a)}}\right]^{2},
\eqno{(2.15)}$$
$$
e^{\bar{\beta}}=(r-a)^{1-\Lambda}(r-b)^{1-\Lambda}
\left[{b(r-a)^{\Lambda}-a(r-b)^{\Lambda}\over{(b-a)}}\right]^{2},
\eqno{(2.16)}$$
$$
\bar{\phi}={c\over{a-b}}\ln{r-a\over{r-b}}=\sqrt{1-\Lambda^{2}\over{2}}\ln\left({r-a\over{r-b}}\right),
\eqno{(2.17)}$$
and
$$
\bar{F_{14}}=qe^{-\bar{\beta}}.
\eqno{(2.18)}$$
The constants $\Lambda, a, b, c$ are related by the equations
$$
2\Lambda^{2}ab=q^{2},
\eqno{(2.19a)}$$
$$
a+b=2m,
\eqno{(2.19b)}$$
$$
\Lambda^{2}c^{2}=(1-\Lambda^{2})(2\Lambda^{2}m^{2}-q^{2}),
\eqno{(2.19c)}$$
where $m$ and $q$ are recognised to be the total mass and charge of the static sphere.
The constant of integration $c$ is actually a sort of scalar charge. The constant 
$\Lambda$ can take values between 0 and 1. For $\Lambda=1$, $c$ is zero and the scalar
field becomes trivial and equations (2.15) and (2.16) yield the usual Reissner
-Nordstrom(RN) solutions. For q=0, one gets the solution for a static sphere
with a minimally coupled scalar field.
\par
The Ricci scalar $R$, for the spacetime given by equations (2.15) and (2.16), 
looks like
$$
R={(a-b)^{2}\over{2(r-a)^{2-\Lambda}(r-b)^{2-\Lambda}}}
{(1-\Lambda^{2})\over{[b(r-a)^{\Lambda}-a(r-b)^{\Lambda}]^{2}}}.
\eqno{(2.20)}$$
\par
This expression shows that the solution is asymptotically flat.
From the metric tensor components, one finds that, $r=a$ and $r=b$ are
coordinate singularities. It is evident from equation (2.20) that if 
$\Lambda\neq{1}$, for both the surfaces $r=a$ and $r=b$ 
the Ricci scalar blows up, whereas when $\Lambda=1$ for both these surfaces,
$$ R=0, \eqno{(2.21)}$$  
and these surfaces act as event horizons to shield the essential singularity 
at $r=0$. Again in this case, i.e for $\Lambda=1$, the scalar field $\bar{\phi}$
becomes trivial and the line element given by equations (2.15) and (2.16) reduces to 
the wellknown Reissner-Nordstrom solution. So, with a scalar meson field, one can 
conclude that the only charged spherical black hole that exists is the RN black hole.
\par
With the help of the transformations (2.5) and (2.7), one can generate the corresponding
solutions for different nonminimally coupled scalar tensor theories, and check 
whether a black hole other than RN one is possible or not. As the transformation
equation suggests it is apperent that any spherical black hole given by action 
(1.1) will be RN one. However, in the next section, we shall discuss some specific 
examples.
\section{Charged black holes with nonminimally \newline coupled Scalar Fields}
\par
{\large{\bf{Case I: Brans-Dicke Theory}}}
\par
If one makes  the choice 
$$f(\phi)=\phi,$$
and 
$$
h(\phi)={\omega\over{\phi}},
\eqno{(3.1)}$$
where $\omega$ is a dimensionless constant parameter, the action (1.1) reduces to the 
one in Brans-Dicke theory. With the help of equation (2.8), one can write 
$$
\bar{\phi}(\phi)=\sqrt{(2\omega+3)}\ln{\phi\over{\phi_{0}}},
\eqno{(3.2)}$$
which, when compared with (2.17), yields
$$
\phi=\phi_{0}\left({r-a\over{r-b}}\right)^{\sqrt{(1-\Lambda^{2})/{2(2\omega+3)}}}.
\eqno{(3.3)}$$
\par
By using the inverse of the transformation equation (2.5) one can now write 
the metric in Brans-Dicke theory as
$$
\phi_{0}ds^{2}=\left({r-b\over{r-a}}\right)^{\sqrt{(1-\Lambda^{2})/2(2\omega+3)}}
[-(r-a)^{\Lambda}(r-b)^{\Lambda}\left({b(r-a)^{\Lambda}-a(r-b)^{\Lambda}\over{(b-a)}}\right)^{-2}dt^{2}$$
$$
\hspace{30mm}+(r-a)^{-\Lambda}(r-b)^{-\Lambda}\left({b(r-a)^{\Lambda}-a(r-b)^{\Lambda}\over{(b-a)}}\right)^{2}dr^{2}$$
$$
\hspace{20mm}+(r-a)^{1-\Lambda}(r-b)^{1-\Lambda}\left({b(r-a)^{\Lambda}-a(r-b)^{\Lambda}\over{(b-a)}}\right)^{2}(d\theta^{2}+\sin^{2}\theta d\phi^{2})],
\eqno{(3.4)}$$
and the Ricci scalar $R=R^{\mu}_{\mu}$ as
$$
R={\omega\over{\phi^{2}}}g^{\mu\nu}\phi_{,\mu}\phi_{,\nu}$$
$$={(1-\Lambda^{2})\omega\phi_{0}(a-b)^{4}\over{2(2\omega+3)}}
{[b(r-a)^{\Lambda}-a(r-b)^{\Lambda}]^{-2}\over{(r-a)^{2-\Lambda-p}(r-b)^{2+p-\Lambda}}},
\eqno{(3.5)}$$
where $$p={\sqrt{(1-\Lambda^{2})\over{2(2\omega+3)}}}.$$
Now one has a singularity at $r=0$. For $\Lambda\neq{1}$, $r=a$ or $r=b$ surfaces 
also have curvature singularities as evident from the metric (3.4) and the expression 
for the Ricci scalar $R$. For $\Lambda=1$, however, $r=a$ and $r=b$ are not 
curvature singularities and act as event horizons. But in this case,
i.e, for $\Lambda=1$, the scalar field becomes trivial, $\phi=\phi_{0}$, and
the theory reduces to the Einstein-Maxwell system.
\par
\vspace{5mm}
{\large{\bf{Case II: Generalised Scalar Tensor Theories}}}
\par
If $\omega=\omega(\phi)$, instead of being a constant, we get a generalisation 
of Brans-Dicke theory. This type of theory had been proposed by 
Bergman$^{12}$, Wagoner$^{13}$, and Nordtvedt$^{5}$. Different choices 
of $\omega$ as a function of $\phi$ can be made depending on the particular physical 
interest. The action (1.1) reduces to that for Nordvedt's theory if
$$f(\phi)=\phi$$ 
and $$
h(\phi)={\omega(\phi)\over{\phi}},\eqno{(3.6)}$$
In what follows, we shall take up two choices of $\omega(\phi)$ in order to verify the 
no scalar hair theorem.
\par
\vspace{5mm}
{\bf{(i) Barker's choice}}
\par
\vspace{3mm}
In this choice$^{14}$,
$$
\omega(\phi)={4-3\phi\over{2(\phi-1)}}.\eqno{(3.7)}$$
Equation (2.8) yields, on integration, the result 
$$
\bar{\phi}=\arctan{\sqrt{\phi-1}-\sqrt{\phi_{0}-1}\over{1+\sqrt{(\phi-1)(\phi_{0}-1)}}},
\eqno{(3.8)}$$
which, in view of equation (2.17) yields 
$$
\phi=\sec^{2}\left[\arctan{\sqrt{\phi_{0}-1}}+\sqrt{1-\Lambda^{2}\over{2}}\ln{r-a\over{r-b}}\right].
\eqno{(3.9)}$$
\par
With the help of equations (2.15), (2.16), (2.5), (2.7) and (3.9), one can write
the metric as
$$
ds^{2}=\cos^{2}\left[\arctan{\sqrt{\phi_{0}-1}}+\sqrt{1-\Lambda^{2}\over{2}}\ln{r-a\over{r-b}}\right]$$
$$
\hspace{20mm}[-(r-a)^{\Lambda}(r-b)^{\Lambda}\left[{b(r-a)^{\Lambda}-a(r-b)^{\Lambda}\over{(b-a)}}\right]^{-2}dt^{2}$$
$$
\hspace{20mm}+(r-a)^{-\Lambda}(r-b)^{-\Lambda}\left[{b(r-a)^{\Lambda}-a(r-b)^{\Lambda}\over{(b-a)}}\right]^{2}dr^{2}$$
$$
\hspace{20mm}+(r-a)^{1-\Lambda}(r-b)^{1-\Lambda}\left[{b(r-a)^{\Lambda}-a(r-b)^{\Lambda}\over{(b-a)}}\right]^{2}(d\theta^{2}+\sin^{2}\theta d\phi^{2})],
\eqno{(3.10)}$$
and the expression for $R$ becomes 
$$
R={\omega\over{\phi^{2}}}g^{\mu\nu}\phi_{,\mu}\phi{,\nu}-{3\over{\phi}}{\omega^{\prime}
\over{2\omega+3}}g^{\mu\nu}\phi_{,\mu}\phi_{,\nu}$$
$$
={4(1-\Lambda^{2})(a-b)^{4}\over{(r-a)^{2-\Lambda}(r-b)^{2-\Lambda}}}
{\sec^{2}[\arctan{\sqrt{\phi_{0}-1}}+\sqrt{1-\Lambda^{2}\over{2}}\ln{r-a\over{r-b}}] 
\over{\{b(r-a)^{\Lambda}-a(r-b)^{\Lambda}\}^{2}}}.
\eqno{(3.11)}$$
\par
\vspace{5mm}
{\bf{(ii) Schwinger's choice}}
\par
\vspace{3mm}
This case is given by$^{15}$
$$2\omega+3={1\over{\alpha\phi}},\eqno{(3.12)}$$
$\alpha$ being a constant. Following the same procedure as in the Brans-Dicke or Barker's theory
we get the solution for the scalar field in Schwinger's theory as
$$\phi=\left[{1\over{\sqrt{\phi_{0}}}}-\sqrt{(1-\Lambda^{2})\alpha\over{8}}\ln{r-a\over{r-b}}\right]^{-2}
\eqno{(3.13)}$$
and the line element as 
$$\hspace{5mm}ds^{2}=
\left[{1\over{\sqrt{\phi_{0}}}}-\sqrt{(1-\Lambda^{2})\alpha\over{8}}\ln{r-a\over{r-b}}\right]^{2}$$
$$
\hspace{5mm}[-(r-a)^{\Lambda}(r-b)^{\Lambda}\left[{b(r-a)^{\Lambda}-a(r-b)^{\Lambda}\over{(b-a)}}\right]^{-2}dt^{2}$$
$$
\hspace{5mm}+(r-a)^{-\Lambda}(r-b)^{-\Lambda}\left[{b(r-a)^{\Lambda}-a(r-b)^{\Lambda}\over{(b-a)}}\right]^{2}dr^{2}$$
$$
\hspace{20mm}+(r-a)^{1-\Lambda}(r-b)^{1-\Lambda}\left[{b(r-a)^{\Lambda}-a(r-b)^{\Lambda}\over{(b-a)}}\right]^{2}(d\theta^{2}+\sin^{2}\theta d\phi^{2})],
\eqno{(3.14)}$$
and the Ricci Scalar $R$ is given by
$$
R={1-\Lambda^{2}\over{4}}{(a-b)^{4}\over{\{(r-a)(r-b)\}^{2-\Lambda}}}\left\{b(r-a)^{\Lambda}-a(r-b)^{\Lambda}\right\}^{-2}$$
$$
\hspace{40mm}\left[{1\over{\sqrt{\phi_{0}}}}-\sqrt{(1-\Lambda^{2})\alpha\over{8}}\ln{r-a\over{r-b}}\right]^{-2}. 
\eqno{(3.15)}$$
In both these subcases, namely (i) Barker's theory and (ii) Schwinger's theory,
the solutions and the Ricci scalar reveal that $r=a$ or $r=b$ are event horizons 
only if $\Lambda=1$. If $\Lambda\neq{1}$, these surfaces are singular by themselves 
and therefore there will be naked singularities and no black holes. Moreover,
for $\Lambda=1$, the solutions in both these cases reduce to the usual RN solution as $\phi$ 
becomes trivial.
\section{Spherical charged black holes in dilaton gravity}
\par
In the weak field limit of the string theory, one obtains Einstein gravity
along with a nonminimally coupled scalar field, called the dilaton field. 
The action in dilaton gravity is
$$
S=\int d^{4}x\sqrt{-g}[e^{-2\phi}R+4e^{-2\phi}g^{\mu\nu}\phi_{,\mu}\phi_{,\nu}-e^{2\alpha\phi}F^{\mu\nu}F_{\mu\nu}],
\eqno{(4.1)}$$
which is different from the action (1.1) as the matter lagrangian (here the
electromagnetic field, given by $L_{em}=-F_{\mu\nu}F^{\mu\nu}$) also is
coupled to the scalar field. In context of (1.1), action (4.1) can be rewritten as 
$$
S=\int d^{4}x\sqrt{-g}[f(\phi)R-h(\phi)g^{\mu\nu}\phi_{,\mu}\phi_{,\nu}-k(\phi)F^{\mu\nu}F_{\mu\nu}],   
\eqno{(4.2)}$$
where $k(\phi)$ takes care of the coupling between the scalar field and the matter field.
This coupling is retained even under the conformal transformation of the form 
(2.5) and (2.7).
The transformed action looks like 
$$
\bar{S}=\int d^{4}x\sqrt{-\bar{g}}[\bar{R}-{1\over{2}}\bar{g}^{\mu\nu}\bar{\phi}_{,\mu}\bar{\phi}_{,\nu}-\bar{k}(\bar{\phi})\bar{F}^{\mu\nu}\bar{F}_{\mu\nu}], 
\eqno{(4.3)}$$
where $\bar{k}(\bar{\phi})=k(\phi(\bar{\phi}))$. The coupling between the scalar
field and the Maxwell field is broken only if $k(\phi)=1$.The static spherically
symmetric solutions for the field equations relevant to the action (4.3) shows that the asymptotic
behaviour of the black hole is characterised by the mass, the electric charge and a dilatonic
charge (ref. 16 and 17).The survival of this dilatonic charge does not jeopardise
the existence of a well defined horizon and hence a black hole. So the no scalar hair
theorem is not applicable in the case of dilaton gravity. In the special case 
when $k(\phi)=1$, however, the scalar field becomes trivial if one insists on the 
existence of an event horizon. These features of dilaton black holes have been discussed
by Garfinkle et al$^{16}$ and Rakhmanov$^{17}$ amongst others. We also refer to the 
work of Horowitz and Strominger$^{18}$ where the existence of blackholes had been discussed
 in higher dimensions in string theory as well as in p-branes.
\section{Conclusion}
In sections 2 and 3 it is shown that the charged static spherically symmetric asymptotically
flat solutions for the field equations lead to black holes only if the scalar field 
becomes trivial both in cases of minimal and nonminimal coupling of the scalar field.
This result crucially depends on the assumption that the coupling function $f(\phi)$
in equation (1.1) is positive. A negative $f(\phi)$,in the weak field limit, however, 
gives rise to a negative Newtonian constant G. The work of Mayo and
Bekenstein includes such couplings also although for a limited kind of $f(\phi)$ 
$[f(\phi)=1-\chi\phi^{2}]$. With a necessarily positive effective Newtonian constant  
 of gravity $[f(\phi)>0]$ the present work enlarges the domain of applicability
 of the scalar no hair theorem to a wide class of scalar tensor theories of gravity.
\par 
An important point to note is that for any dimension n other than $n=4$,
a non trivial coupling between the scalar field and the Maxwell field 
exist in the transformed action after the conformal transformation
in different general scalar tensor theories also and thus for higher ( or lower)
dimensional theories the black holes are different from black holes having no scalar hair 
in four dimension. This difference between $n=4$ and $n\neq{4}$ arises because
equation (2.10) is valid only for $n=4$. A recent work of Cai and Myung$^{19}$ in
Brans-Dicke theory can be referred to in this context. 
\par
These results consolidates the no scalar hair theorem, proved earlier by Bekenstein
and Mayo-Bekenstein in a different approach for a charged spherical black hole
in four dimensions for a positive definite $f(\phi)$ 
when the electromagnetic field itself is not coupled to the scalar field.
\par
It remains to be seen, however, whether this remains valid along with scalar  
potentials in an arbitrary scalar tensor theory. The case of a negative $f(\phi)$
 also deserves extensive investigation in a wider class of scalar tensor theories 
 particularly in view of the results obtained by Bekenstein$^{7}$. It will also be   
 important to investigate the stability of the black hole solution, if any, in such cases. 
For a discussion in detail on this point we refer to the work of Mayo and Bekenstein$^{8}$.\\
\vspace{3mm}\\
{\Large{\bf{Acknowledgement}}}\\
\vspace{1.5mm}\\
One of the authors (S.S) is thankful to University Grants' Commission of India 
for the financial support.
\newpage
{\Large{\bf{References}}}\\
\vspace{5mm}\\
\large{
$^{1}$A.Saa, {\it{J.Math.Phys}}, {\bf{37}}, 2346, (1996)\\
$^{2}$B,C.Xanthopoulos and T.Zannias, {\it{Phys.Rev.D}}, {\bf{40}}, 2564, (1989)\\
$^{3}$J.D.Bekenstein, {\it{Ann.Phys.}}, {\bf{82}}, 535, (1974) \\
$^{4}$C.Brans and R.H.Dicke, {\it{Phys.Rev}}, {\bf{124}}, 925, (1961)\\
$^{5}$K.Nordtvedt Jr. {\it{Astrophys.J}}, {\bf{161}}, 1059, (1970)\\
$^{6}$R.Penney, {\it{Phys.Rev}}, {\bf{182}},1383, (1969)\\
$^{7}$J.D.Bekenstein, {\it{Ann.Phys.}}, {\bf{91}}, 75 (1975)\\
$^{8}$A.E.Mayo and J.D.Bekenstein, {\it{Phys.RevD}}, {\bf{54}}, 5059, (1996)\\
$^{9}$J.D.Bekenstein, {\it{Phys.Rev.D}}, {\bf{51}}, R6608, (1995)\\
$^{10}$R.H.Dicke, {\it{Phys.Rev}}, {\bf{125}}, 2163, (1962) \\
$^{11}$J.L.Synge, {\bf{Relativity: The General Theory}} (North Holland Publishing Co. (1960))\\   
$^{12}$P.G.Bergman, {\it{Int.J.Theor.Phys.}}, {\bf{1}}, 25 (1968)\\
$^{13}$R.V.Wagoner, {\it{Phys.Rev.D}}, {\bf{1}}, 3209, (1970)\\
$^{14}$B.M.Barker, {\it{Astrophys.J.}}, {\bf{219}}, 5, (1978) \\
$^{15}$J.Schwinger, {\bf{Particles, Sources and Fields}} (Addison-Wesley, Reading,Massachusetts, (1970))\\
$^{16}$D.Garfinkle, G.T.Horowitz and A.Strominger, {\it{Phys.Rev.D}}, {\bf{43}}, 3140 (1991)\\
$^{17}$M.Rakhmanov, {\it{Phys.RevD}}, {\bf{50}}, 5155, (1994)\\
$^{18}$G.T.Horowitz and A.Strominger, {\it{Nucl.Phys.B}}, {\bf{360}}, 197, (1991)\\
$^{19}$R.G.Cai and Y.S.Myung, {\it{Phys.Rev.D}}, {\bf{56}}, 3466, (1997)\\}
\end{document}